# Data Stewardship Decoded

## *Mapping Its Diverse Manifestations and Emerging Relevance at a time of AI*

**Dr. Stefaan G. Verhulst,**

**Co-Founder The GovLab and The DataTank**

**Research Professor Tandon School of Engineering, New York University**

## Abstract

*Data stewardship has become a critical component of modern data governance, especially with the growing use of artificial intelligence (AI). Despite its increasing importance, the concept of data stewardship remains ambiguous and varies in its application. This paper explores four distinct manifestations of data stewardship to clarify its emerging position in the data governance landscape. These manifestations include a) data stewardship as a set of competencies and skills, b) a function or role within organizations, c) an intermediary organization facilitating collaborations, and d) a set of guiding principles.*

*The paper subsequently outlines the core competencies required for effective data stewardship, explains the distinction between data stewards and Chief Data Officers (CDOs), and details the intermediary role of stewards in bridging gaps between data holders and external stakeholders. It also explores key principles aligned with the FAIR framework (Findable, Accessible, Interoperable, Reusable) and introduces the emerging principle of AI readiness to ensure data meets the ethical and technical requirements of AI systems.*

*The paper emphasizes the importance of data stewardship in enhancing data collaboration, fostering public value, and managing data reuse responsibly, particularly in the era of AI. It concludes by identifying challenges and opportunities for advancing data stewardship, including the need for standardized definitions, capacity building efforts, and the creation of a professional association for data stewardship.*

# Introduction

Data has emerged in recent years as an increasingly vital component of society and everyday life. This trend, the so-called datafication of the world, is only likely to intensify with the advent of artificial intelligence (AI).

Data is the fuel that underpins the rapid strides made in generative, predictive and other forms of AI.

Yet with growing ubiquity come greater challenges, or at least questions. If data is the foundation for much contemporary human existence, then it is increasingly important to address how that data is made accessible and managed. Addressing the risks and tradeoffs—and opportunities—surrounding data governance has become one of the most pressing public policy tasks of our times, with important decisions to be made at every link in the data value chain.

Data governance is a multi-threaded challenge; solutions and appropriate governance frameworks will include multiple steps, mechanisms, and institutions. In what follows, we focus on one particularly important aspect of the data governance landscape: the role and functions of data stewardship.

Data stewardship is a term increasingly used in the context of modern data governance, potentially offering solutions or mitigating mechanisms for better management of digital assets. Yet it is also a term that is fraught with ambiguity, its definition and scope often contested. For data stewardship to evolve into a viable solution, we need a much better understanding of what it means, and be more specific when using the concept.

This essay aims to provide clarity by exploring different manifestations of data stewards and the notion of data stewardship. In particular, we aim to advance our understanding of data stewardship by examining four manifestations of the notion:

1. a set of competencies and skills;
2. a function or role within an organization;
3. an intermediary organization; and
4. a set of principles.

After discussing these various manifestations, we focus on the intersection of data stewardship with AI. Finally, we conclude by exploring some remaining challenges and areas for further exploration. Our overall goal is to provide a conceptual, relevant, and practical framework to help understand and advance the functions of data stewards in a twenty-first-century datafied era.

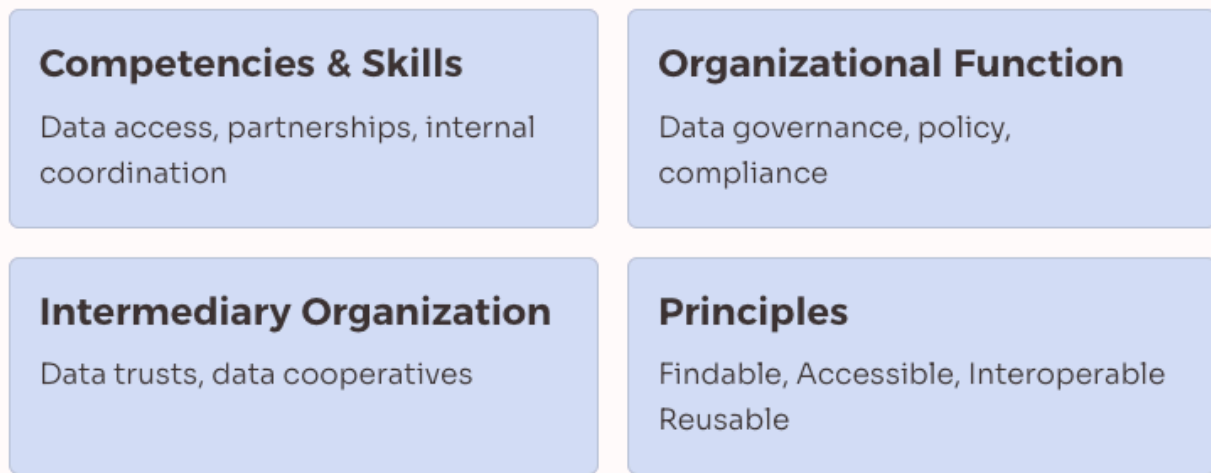

Figure 1: Data Stewardship Manifestations

## I.     Data Stewardship as a Set of Competencies and Skills

Data stewardship is a broad concept that encompasses many competencies and skills. At a high level, the responsibility of a data steward is to ensure end-to-end responsibility throughout the data value chain—collection, storage, handling, use, and reuse of data. In addition, one of the central goals of data stewardship is to break down data silos (e.g., in privately held datasets) and create public value by fostering inter-sectoral collaborations and reusing data and data expertise.

These high-level goals and principles can be broken down into five more specific competencies and skills:

1. **Stewarding Data: Data Audit, Assessment & Governance**: Stewarding data assets for the public interest involves formulating and determining priority questions, scoping and iterating on assessments of "minimum viable" data needed, identifying and documenting data assets, and considering ethical and fundamental rights implications. It also includes establishing operational, technical, and governance models to measure impact.
2. **Stewarding Relationships: Partnership and Community Engagement**: This competency involves stewarding relationships by reaching out to and vetting potential partners while informing beneficiaries of insights generated from data initiatives. Data stewards act as points of contact for data re-use, engage stakeholders in user-driven processes, and help establish a social license for data re-use through community deliberations and agreements.
3. **Stewarding Internal Resources: Internal Coordination and Data Operations**: Data stewards must gain approval and coordinate with internal actors, ensuring all stakeholders are informed and aligned. They establish data operations to map and match internal resources, expertise, and skills needed to enable data collaboration.
4. **Stewarding Data Collaboration: Nurturing Data Collaboratives for Sustainability**: Data stewards work to institutionalize data innovation by making data re-use systematic. This involves developing business cases for scaling and sustaining data innovation, measuring impact, and sharing insights to build societal and business cases for collaboration.
5. **Stewarding Insights: Dissemination and Communication of Findings**: Acting as the face of data projects, data stewards are responsible for raising awareness about shared outcomes with users, partners, and governments. They translate [data intelligence into actionable decision intelligence](#) and ensure regulatory compliance and contractual obligations are communicated effectively.

## II. The Function or Role of Data Stewards Within Organizations

What do these various principles and competencies mean in practice? For organizations considering appointing data stewards, questions remain about how they fit into overall organizational charts, what their specific responsibilities may be, and how they might interact with other groups or individuals that already handle data. In particular, there may be some ambiguity about potential overlap between the roles of existing Chief Data Officers (CDOs) and data stewards. Below, we seek to clarify some of these points of potential confusion by providing a roadmap for companies and other entities considering appointing data stewards.

Currently, individuals or groups in organizations tasked with managing data typically have four key responsibilities:

1. **Data Custodian**: Managing data storage and security.
2. **Compliance Officer**: Ensuring adherence to legal and ethical standards.
3. **Data Strategist**: Aligning data practices with organizational goals.
4. **Data Facilitator**: Bridging technical and non-technical teams.

These roles are not necessarily assumed by data stewards (although they may come into contact with some of these tasks). Instead, the responsibilities may continue to be assigned to existing individuals or entities, most commonly Chief Data Officers. The GovLab has advocated for a more [strategic role for Data Stewards,](#) one that would complement the functions of CDOs. Both positions are integral to strategic data management and governance within organizations, but they serve distinct functions:

- **Chief Data Officers (CDOs)** are primarily responsible for the overarching governance and [utilization of data as a strategic asset within organizations](#). Their duties often encompass data management, privacy, security, and ensuring compliance with relevant regulations. The role is typically inward-facing, focused on optimizing data processes and policies to enhance organizational efficiency and decision-making.

- **Data Stewards** are proactive agents who facilitate data access and collaboration, particularly across sectors, with a key goal of generating public value. Their responsibilities typically include collaborating with external partners to unlock the value of data for public interest initiatives; protecting stakeholders from potential harms associated with data re-use; ensuring that insights derived from data are effectively applied to address societal challenges.

Three key differences between the functions of Data Stewards and CDOs are worth highlighting:

1. **Scope of Responsibilities**: While CDOs concentrate on internal data governance, Data Stewards extend their focus to include external data collaborations aimed at societal benefits.
2. **Orientation**: CDOs are inward-focused, enhancing internal data processes; Data Stewards are outward-focused, promoting data sharing and collaboration beyond organizational boundaries.
3. **Skill Set**: Data Stewards require skills in partnership building, ethical data management, and the ability to operationalize data insights for public good, which may extend beyond the traditional purview of CDOs.

The GovLab underscores the importance of both roles (and that of the Chief Data Protection Officer) in the modern data ecosystem, advocating for a clear delineation of responsibilities to ensure that data is managed effectively within organizations and leveraged collaboratively to address broader societal issues. In effect, data stewards and CDOs must work together to bring greater responsibility, transparency and effectiveness to how data assets are deployed.

### III. Data Stewardship as an Intermediary Function

Among the most important functions of data stewards is to build and maintain collaborations and partnerships. As such, one of the key manifestations of data stewardship is that of an intermediary function. Data stewards perform this function between and among multiple stakeholders in the data ecology—data holders and users, private sector data "owners" and third-party public or civil society groups seeking to repurpose the data, and more. These [data collaboratives](#) can take different forms including [data trusts](#), [data cooperatives](#), [data altruistic organizations](#) or [data commons](#). Among the chief desired outcomes of this function is to enable *privately held* data to be responsibly and ethically repurposed for the *public* good.

The intermediary function can be broken down into three roles:

1. **Trusted Intermediary**: Acting as a neutral entity that ensures data is (re)used responsibly and equitably.
2. **Negotiator**: Facilitating [digital self determination](#) and the creation of agreements on data reuse, such as access rights and usage limitations.

3. **Facilitator**: Providing tools, resources, and frameworks to simplify data sharing and collaboration.

Several organizations have drawn attention—in different ways--to the intermediary function that can be performed by data stewards. The Aapti Institute, for instance, emphasizes the need for [participatory and equitable data practices as a core element of data stewardship](). Highlighting the importance of community involvement in data governance, this group advocates for an intermediary function that empowers individuals and groups, ensuring that data governance models are inclusive and just.

Similarly, the Ada Lovelace Institute focuses on [participatory data stewardship](), and specifically on involving populations in the governance of their data to ensure that data practices are transparent, equitable, and beneficial to society. This group proposes a framework that includes various different types of public participation, including informing, consulting and collaborating, and empowering individuals in data governance processes. By seeking to rebalance power dynamics in data governance, the Ada Lovelace Institute emphasizes the importance of ensuring that data use and reuse align with societal values and public interest.

## IV.    The Principles of Data Stewardship

Finally, in considering the various manifestations of data stewardship, it is important to consider some of the underlying principles that guide its approach to data and data handling. At a broad level, the principles of data stewardship align closely with the well-established [FAIR principles](): Findable, Accessible, Interoperable, and Reusable. In addition, we suggest that twenty-first century notions of data stewardship add a further principle of AI Readiness.

Below, we offer a breakdown of how data stewardship complements these principles:

1. **Findable**:
    - Ensure that data is well-documented and indexed so that users can locate it easily.
    - Develop and maintain metadata standards to describe data comprehensively.
    - Assign unique and persistent identifiers (e.g., DOIs) to datasets.
2. **Accessible**:
    - Facilitate access to data while adhering to legal, ethical, and regulatory requirements.
    - Implement clear and secure access protocols, including user authentication and authorization processes.
    - Define access conditions, such as open data licenses or agreements for restricted access.
3. **Interoperable**:
    - Promote the use of standardized formats, vocabularies, and ontologies to enable seamless integration and reuse of data.

- Advocate for cross-sectoral and international compatibility in data systems and standards.
- Ensure data is machine-readable and aligned with semantic frameworks.

4. **Reusable**:
    - Ensure data quality and relevance by maintaining accurate, complete, and updated datasets.
    - Apply clear licensing and data usage policies to define conditions for reuse.
    - Support long-term preservation and curation of data to maintain its usability over time.

5. **AI Ready:**
    - Ensure [datasets](#) are annotated, labeled and structured to support AI training and testing.
    - Establish mechanisms for identifying and mitigating bias within datasets- from the collection to the analysis of the data- to improve fairness and representativeness.
    - Incorporate explainability standards to align data with transparency requirements and general ethical AI principles.

| Manifestation | Description | Key Components |
|---|---|---|
| **Set of Competencies and Skills** | Represents the abilities required to manage data across its lifecycle and foster collaboration for public value. | - Data audit, assessment, and governance<br><br>- Partnership and community engagement<br><br>- Internal coordination and data operations<br><br>- Sustaining data collaboration<br><br>- Communicating insights |

| Function or Role | Defines specific responsibilities within organizations, focusing on bridging technical and strategic needs. | - Complementing Chief Data Officers (CDOs)<br><br>- Facilitating external data access and collaborations<br><br>- Proactive partnership building<br><br>- Ethical and impactful data application |
|---|---|---|
| Intermediary Organization | Acts as a bridge between diverse stakeholders, enabling responsible and ethical data sharing for public good. | - Trusted intermediary ensuring fairness<br><br>- Negotiator of access and usage agreements<br><br>- Facilitator providing frameworks and tools for collaboration |
| Set of Principles | Emphasizes values that guide responsible data stewardship, integrating technical and ethical dimensions. | - FAIR principles: Findable, Accessible, Interoperable, Reusable<br><br>- AI Readiness: Bias mitigation, transparency, and explainability |

Figure 2: Summary of Data Stewardship Manifestations.

## V. Data Stewardship in the Era of Artificial Intelligence (AI)

As noted, the rise of AI has both changed some of the expected functions of Data Stewards and elevated their importance to unprecedented levels. Modern AI systems are heavily reliant on high-quality datasets to function effectively. In addition, incomplete or unrepresentative data can lead to biases in AI systems that may create new or exacerbate existing social and economic divisions and forms of marginalization. For all these reasons, data governance provides the bedrock of AI governance, and (chief) data stewards play a key role in the age of AI.

Some particular considerations would include (see also above in making data AI Ready):

1. **Ensuring Data Quality**: AI requires accurate, comprehensive, and representative datasets. Data stewards play a critical role in curating, validating, and maintaining these datasets to minimize biases and errors.
2. **Responsible Data Reuse**: As AI applications often involve sensitive and personal data, data stewards ensure compliance with AI regulations and ethical standards, mitigating risks associated with misuse or unintended consequences.
3. **Facilitating Cross-Sector Collaboration**: AI development often relies on diverse data sources. Data stewards act as intermediaries to enable the creation of data commons or data sharing across sectors while maintaining trust and safeguarding data integrity.
4. **Transparency and Accountability**: Data stewards can promote transparency in how AI models use data, ensuring stakeholders understand data provenance, processing methods, and potential biases.
5. **Supporting AI Innovation**: By establishing robust data governance frameworks, data stewards can create environments where AI innovations can thrive, balancing innovation with responsibility. They reduce the risk of *missed*-use, even while protecting against misuse.

The precise implications of many of these functions remain to be determined. We are still in the early stages of the AI era, and new challenges and opportunities are likely to present themselves, requiring a dynamic approach to how we define the role of data stewards. Nonetheless, it is already clear that the integration of AI into various domains underscores the need for a proactive approach to data stewardship, where data is not only managed responsibly but also optimized to maximize its potential for societal benefit.

## VI.    Looking Forward: Challenges and Opportunities

Data stewardship is a multifaceted concept encompassing competencies, roles, intermediary functions, and principles. Its diverse manifestations and dimensions highlight the complexity of managing data responsibly and effectively, especially in the age of AI. By understanding these dimensions and their interconnections, organizations can overcome confusion and implement stewardship practices that maximize data's value while safeguarding ethical standards.

As with so much of the digital world, the landscape for data stewardship going forward is characterized by both risk and opportunity. Often, these are two sides of the same coin: well-managed challenges can unleash new potential. In closing, we highlight three risks, and then follow with three pathways to transform those risks into opportunity.

**Three Risks:**

1. **Ambiguity**: The broad and multi-faceted nature of data stewardship may create confusion, making it difficult for organizations to implement effective practices and leading to under-utilized or irresponsibly managed data assets.
2. **Resistance to Change**: Adopting stewardship practices and principles, especially those that require making data accessible in a more systematic manner, may encounter institutional inertia or skepticism.

3. **Resource Constraints**: Effective stewardship requires investment in skills, technologies, and infrastructure. The case for these investments may be hard to make, especially given the ambiguity.

**From Risk to Opportunity:**

1. **Standardize Definitions**: Clarify the scope of data stewardship to align expectations and practices. This can lead to a more widespread adoption and implementation of data stewardship practices, helping unleash the potential of under-utilized data assets.
2. **Promote Capacity Building**: Train individuals in the competencies required for stewardship roles. This can lead to more systematic, sustainable and responsible data re-use.[1]
3. **Foster Collaboration and Change Management**: Leverage the intermediary function of data stewardship to build trust, establish [digital self-determination](#), and drive innovation. Collaboration is one of the most important change management objectives needed toward driving innovation and inclusion in the twenty-first century.

**The Need for a Professional Association**

Many of these risks could be managed—and potential unleashed—with more formal structures dedicated to data stewardship. In conclusion, we therefore suggest establishing a professional society or association dedicated to data stewardship. This group, which would be multi-sectoral and international in nature, would help standardize practices, advocate for resources, and support professional development of data stewards. Such an association could also play a crucial role in addressing the challenges and fostering a global community of practitioners committed to advancing responsible data governance. In so doing, it would help fulfill the tremendous potential of this critical twenty-first century asset in a manner that was responsible, inclusive, and innovation-enhancing.

---

[1] The Govlab and The Data Tank have trained data stewards across sectors and around the globe See: https://datatank.org/data-stewards-bootcamp/ & https://course.opendatapolicylab.org/